\documentclass[12pt,a4paper,notitlepage]{article}
\usepackage{graphics}
\usepackage{amsmath}
\usepackage{graphicx}
\usepackage{caption}
\usepackage{cite}
\title{Charmonium Properties Using the Discrete Variable Representation (DVR) Method}
\author{Bhaghyesh
\\
\\
Department of Physics, Manipal Institute of Technology
\\
Manipal Academy of Higher Education, Manipal, India 576104
\\
email: bhaghyesh.mit@manipal.edu
\\
ORCID: 0000-0003-3994-9945}
\date{}
\begin{document}
\maketitle
\begin{abstract}
The Schr\"odinger equation is solved numerically for charmonium using the discrete variable representation (DVR) method. The Hamiltonian matrix is constructed and diagonalized to obtain the eigenvalues and eigenfunctions. Using these eigenvalues and eigenfunctions, spectra and various decay widths are calculated. The obtained results are in good agreement with other numerical methods and with experiments.
\end{abstract}
%\clearpage
\section{Introduction}
The first quarkonium state was discovered independently at SLAC \cite{slac} and BNL \cite{bnl}, confirming the existence of heavy quark bound states. Since then, quarkonium have always been of great interest to particle physicists, being one of the extensively investigated system both theoretically and experimentally \cite{QWG,snowmass}. New states are continuosly being detetcted at various experiments. Recently the LHCb collaboration \cite{lhcb} has detected a new state $X(3842)$ which is interpreted as a candidate for the unobserved $\psi_3(1~^3D_3)$ state. Both charmonium and bottomonium have rich spectrum of states below the open flavor threshold which have been experimentally observed and various decay widths of these states have also been measured \cite{pdg}. Studies on heavy quark systems are important because it gives information about quark interaction potential, confinement, QCD coupling constant, CKM matrix elements, and various other inputs to the standard model, some of which cannot be directly obtained from QCD.
\\
Theoretically, quarkonium systems have been studied by various formalisms based on phenomenological potential models \cite{lucha1,godfrey,barnes,qing,olga}, effective field theory \cite{eft}, lattice gauge theory \cite{lattice1,lattice2,lattice3,lattice4}, Bethe Salpeter equation \cite{bs1,bs2,bs3,bs4}, etc. Among these, owing to its simplicity, formalism based on potential models is the widely chosen method to investigate quarkonium systems. In this method, both relativistic and quantum corrections can be easily incorporated.  Potential models have been highly successful in predicting the spectra and decay widths \cite{barnes,qing,olga}. In potential models, the usual method is to extract the properties of quarkonium by solving the Schr\"odinger equation using a chosen quark-antiquark potential. The widely used quark-antiquark potential in phenomenological models is the so called Cornell potential \cite{corn1,corn2,corn3,corn4,origin}, which includes a short range Coulomb term and a linear confinement term. The form of this  potential is also confirmed by lattice QCD calculations \cite{bali1,bali2}.
\\
The Schr\"odinger equation for most of the $q \bar{q}$ potentials (including, the Cornell potential) cannot be solved analytically; hence numerical solutions are called for. Some of the methods found in literature for solving the Schr\"odinger equation for $q\bar{q}$ systems are: numerical methods based on Runge-Kutte approximation \cite{lucha_math,vinod}, Numerov matrix method \cite{numerov1,numerov2,numerov3}, asymptotic iteration method \cite{aim1,aim2,aim3}, Fourier grid Hamiltonian method \cite{fgh}, variational method \cite{variational1,variational2}, etc. Another method for numerically solving the Schr\"odinger equation is the discrete variable representation (DVR) method. This method has not been applied to quarkonium spectroscopy. Hence, in this article, we numerically solve the Schr\"odinger equation for $c\bar{c}$ system  using the discrete variable representation (DVR) scheme of Colbert and Miller \cite{colbert}. DVR method was initially introduced by Harris \cite{harris}, and was extensively developed by Light and co-workers \cite{light1,light2,light3,light4,light5,light6}. DVR's provide highly efficient and accurate solutions to quantum dynamical problems and have been widely used in atomic physics and quantum chemistry \cite{app1,app2,app3,app4,app5,app6,app7,app8,app9,app10}. More details on DVR methods can be found in refs.\cite{light,dvr_ref}.
\\
This paper is organised as follows: a brief discussion on the potential model used to describe the $c\bar{c}$ system and the DVR scheme used to solve the Schr\"odinger equation are presented in Section \ref{sec2}. The various decay properties calculated in the present analysis are given in Section \ref{sec3}. Results and discussions of the present work are given in Section \ref{sec4}.

\section{Formalism}\label{sec2}
\begin{figure}[h]
	\begin{minipage}{.48\textwidth}
		\centering
		% include first image
		\includegraphics[width=1.\textwidth]{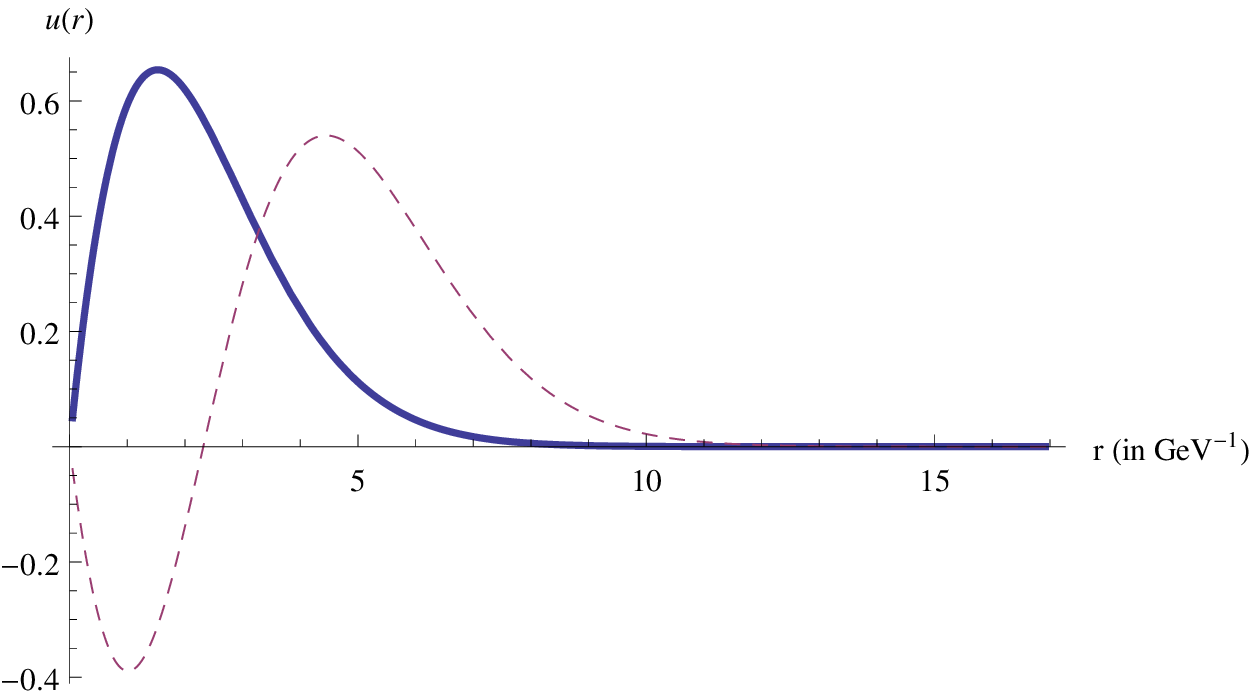}  
		%\caption{}
		%\label{fig:sub-first}
	\end{minipage}\hfill
	\begin{minipage}{.48\textwidth}
		\centering
		% include second image
		\includegraphics[width=1.\textwidth]{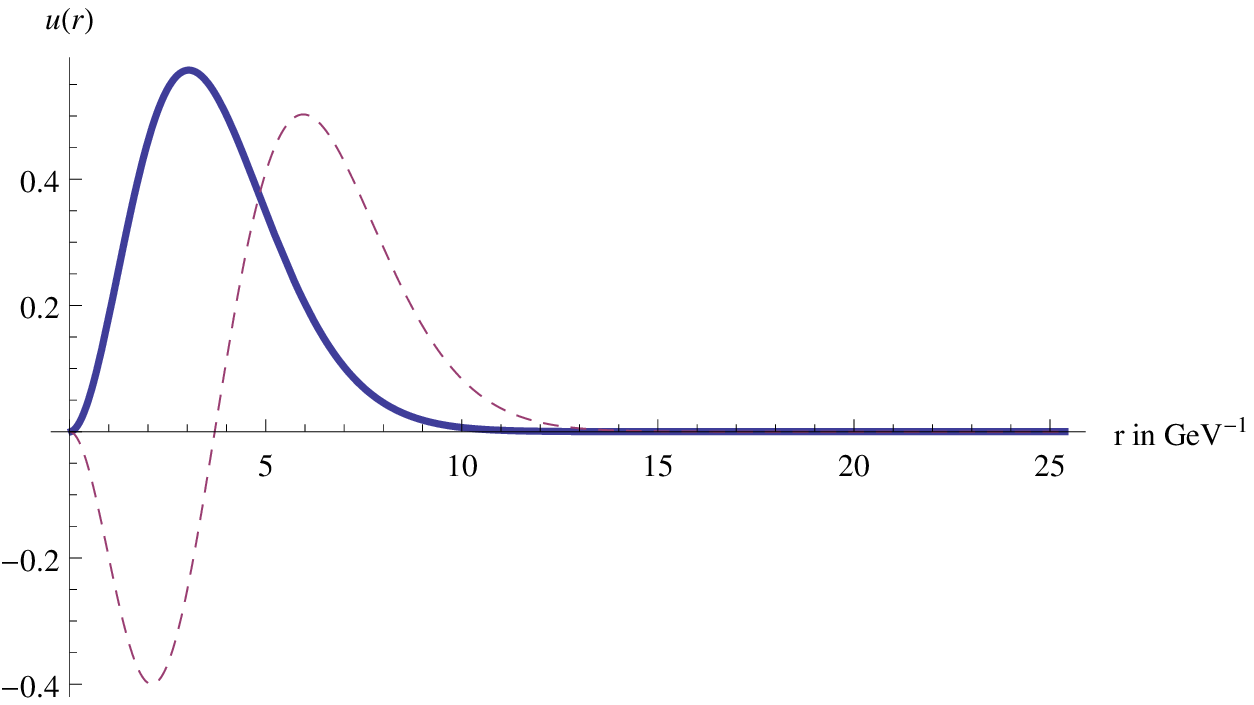}  
		%\caption{}
		%\label{fig:sub-second}
	\end{minipage}
	\caption{The wavefunctions of charmonium $^3S_1$ states (left) and $^3P_J$ states (right). Solid curve represent the $n=1$ state while dashed curve represent the $n=2$ state.}
	\label{fig}
\end{figure}

As a minimal model describing charmonium, we have used a nonrelativistic potential model, with the Hamiltonian
\begin{equation}
H = M + \frac{p^2}{2\mu} + V(r),\label{hamil}
\end{equation}
where $p$ is the relative momentum, $\mu~(= m_c m_{\bar{c}}/m_c+m_{\bar{c}})$ is the reduced mass of the $c\bar{c}$ system, $M = m_c+m_{\bar{c}}$, and $V(r)$ is the quark-antiquark potential. $m_c$ and $m_{\bar{c}}$ are the masses of individual quark and antiquark, respectively.
For describing the quark-antiquark interaction, we use the standard Cornell potential plus  a Gaussian-smeared contact hyperfine interaction \cite{barnes}:
\begin{equation}
V(r) = -\frac{4}{3}\frac{\alpha_c}{r} + br + V_0 +\frac{32\pi\alpha_c}{9m_c^2}\left(\frac{\sigma}{\sqrt{\pi}}\right)^3 e^{-\sigma^2 r^2} \vec{S}_c\cdot \vec{S}_{\bar{c}},\label{pot}
\end{equation}
Parameters used in eq.(\ref{pot}) are given in Table (a) and are obtained by fitting the spectrum. Charmonium properties can be obtained by solving the Schr\"odinger equation corresponding to the Hamiltonian given in eq.(\ref{hamil}) with potential given in eq.(\ref{pot}). In this work, to solve the Schr\"odinger equation we have used the DVR scheme of Colbert and Miller \cite{colbert}. In the DVR, the Hamiltonian is represented by a matrix on a uniform grid of points ($r_i=i\Delta r, i = 1,2,3,..$) in the coordinate space. Once the H-matrix is constructed, diagonalization gives us the bound state eigenvalues and the amplitudes of eigenfunctions on the grid point chosen. 
\\
In ref.\cite{colbert} authors have shown that the kinetic energy matrix can be written as
\begin{align}
T_{ij} = \frac{\hbar^2}{2m\Delta r^2} (-1)^{i-j}\left\{ \begin{array}{cc} 
\pi^2/3-1/2i^2 & \hspace{5mm} i=j \\
\frac{2}{(i-j)^2}-\frac{2}{(i+j)^2} & \hspace{5mm} i \neq j \\
\end{array} \right.\label{ke}
\end{align}
with $r_i = i\Delta r, (i = 1,2,...)$, where $\Delta r$ is the grid spacing. The potential energy matrix is diagonal
\begin{equation}
V_{ij}=V(r_i)\delta_{ij} ~.\label{pe}
\end{equation}
\begin{table}
	\centering
	\caption{Parameters used in the model}
	\label{para}       % Give a unique label
	%For LaTeX tables use
	\begin{tabular}{ll}
		\hline\noalign{\smallskip}
		Parameter & Value \\
		\noalign{\smallskip}\hline\noalign{\smallskip}
		$\alpha_c$     & 0.54 \\
		$b$     & 0.136 GeV$^2$ \\
		$V_0$    & 0.149 GeV \\
		$\sigma$ & 1.1 GeV  \\
		$m_c$    & 1.4 GeV \\   	
		\noalign{\smallskip}\hline
	\end{tabular}
\end{table}
\noindent  We have used eqs.(\ref{hamil},\ref{pot},\ref{ke},\ref{pe}) to construct the Hamiltonian matrix in the present model, which upon diagonalization returns the bound state eigenvalues and the amplitudes of eigenfunctions on the chosen grid points. In the present anlysis we have chosen a grid of length 10 fm with 1000 grid points. The eigenvalue problem for the matrix of the Hamiltonian (\ref{hamil}) was solved using Mathematica. For a given eigenvalue, in order to obtain the eigenfunction in the entire range of coordinates, we have used the built-in interpolation function in Mathematica through the obtained eigenfunctions on grid points. This interpolation function was used as the representation of the reduced radial wavefunction for our further analysis. Obtained wavefunctions for $^3S_1$ and $^3P_J$ states are shown in Figure \ref{fig}.  
In order to compute fine structure of the $L \neq 0$ states, we add the spin-orbit and tensor terms perturbatively \cite{barnes}:
\begin{equation*}
V_{fs} = \frac{1}{m_c^2}\left[ \left( \frac{2\alpha_c}{r^3} - \frac{b}{2r} \right) \vec{L}\cdot\vec{S} + \frac{4 \alpha_c}{r^3} T \right]
\end{equation*}
The computed mass spectra of charmonium are listed in Table \ref{mass}. Using the obtained wavefunction we also compute the root mean square radii ($\sqrt{<r^2>}$) and the square of the radial wavefunction at the origin ($|R(0)|^2$) for these states and our results are listed in Table \ref{radii}.
\begin{table}
	\centering
	\caption{Mass spectrum of charmonium states (in MeV)}
	\label{mass}       % Give a unique label
	%For LaTeX tables use
	\begin{tabular}{lllllllll}
		\hline
		\noalign{\smallskip}
		State                              & Present & EXP \cite{pdg}       & \cite{barnes} & \cite{qing} & \cite{akrai} & \cite{soni} & \cite{akbar1} & \cite{debastiani} \\
		\noalign{\smallskip}\hline\noalign{\smallskip}
		$J/\psi$ 	 & 3097    & 3096.900$\pm$0.006   & 3090    & 3097 	& 3094  & 3094        & 3090.0	 	& 3091.7   \\
		$\eta_c$(1S) & 2986    & 2983.9 $\pm$ 0.5     & 2982    & 2979  & 2995  & 2989        & 2981.6      & 2992.4   \\
		$\psi$(2S)   & 3663    & 3686.097 $\pm$ 0.025 & 3672    & 3673  & 3649  & 3681        & 3671.8      & 3671.4   \\
		$\eta_c$(2S) & 3620    & 3637.6 $\pm$ 1.2     & 3630    & 3623  & 3606  & 3602        & 3630.3      & 3631.7   \\
		$\psi$(3S)   & 4055    & 4039 $\pm$ 1         & 4072    & 4022  & 4036  & 4129        & 4071.6      & 4075.5   \\
		$\eta_c$(3S) & 4025    &                      & 4043    & 3991  & 4000  & 4058        & 4043.2      & 4048.1   \\
		$\psi$(4S)   & 4385    & 4421 $\pm$ 4         & 4406    & 4273  & 4362  & 4514        & 4406.1      & 4415.0   \\
		$\eta_c$(4S) & 4360    &                      & 4384    & 4250  & 4328  & 4448        & 4383.7      & 4393.3   \\
		$\psi$(5S)   & 4678    & 			          &         & 4463  & 4654  & 4863        & 4703.8      &        \\
		$\eta_c$(5S) & 4657    &                      &      	& 4446  & 4622  & 4799        & 4685.0      &        \\
		$\psi$(6S)   & 4947    &                      &         & 4608  & 4925  & 5185        & 4976.9      &        \\
		$\eta_c$(6S) & 4929    &                      &         & 4595  & 4893  & 5124        & 4960.4      &        \\
		$\chi_{c2}$(1P)& 3546  & 3556.17 $\pm$ 0.07   & 3556    & 3554  & 3556  & 3480        & 3549.0      & 3548.1   \\
		$\chi_{c1}$(1P)& 3498  & 3510.67 $\pm$ 0.05   & 3505    & 3510  & 3523  & 3468        & 3505.4      & 3501.8   \\
		$\chi_{c0}$(1P)& 3419  & 3414.71 $\pm$ 0.30   & 3424    & 3433  & 3457  & 3428        & 3424.5      & 3425.8   \\
		$h_c$(1P)   & 3507     & 3525.38 $\pm$ 0.11   & 3516    & 3519  & 3534  & 3470        & 3515.6      & 3510.5   \\
		$\chi_{c2}$(2P)& 3955  & 				      & 3972    & 3937  & 3956  & 3955        & 3964.8      & 3970.0   \\
		$\chi_{c1}$(2P)& 3909  & 					  & 3925    & 3901  & 3925  & 3938        & 3924.9      & 3925.8  \\
		$\chi_{c0}$(2P)& 3839  & 				      & 3852    & 3842  & 3866  & 3897        & 3852.3      & 3856.7  \\
		$h_c$(2P)    & 3918    & 				      & 3934    & 3908  & 3936  & 3943        & 3933.6      & 3933.4  \\
		$\psi_3$(1 $^3D_3$)& 3791&                    & 3806    & 3799  & 3801  & 3755        & 3805.3      & 3800.6  \\
		$\psi_2$(1 $^3D_2$)& 3786&                    & 3800    & 3798  & 3805  & 3772        & 3800.4      & 3796.7  \\
		$\psi$(1 $^3D_1$)  & 3771& 3773.13 $\pm$ 0.35 & 3785    & 3787  & 3799  & 3775        & 3785.0      & 3783.1  \\
		$\eta_{c2}$(1 $^1D_2$)& 3785&                 & 3799    & 3796  & 3802  & 3765        & 3799.4      & 3795.1   \\
		$\psi_3$(2 $^3D_3$)& 4146&                    & 4167    & 4103  & 4151  & 4176        & 4165.5      & 4167.1   \\
		$\psi_2$(2 $^3D_2$)& 4138&                    & 4158    & 4100  & 4152  & 4188        & 4158.2      & 4160.2   \\
		$\psi$(2 $^3D_1$)  & 4122& 4191 $\pm$ 5       & 4142    & 4089  & 4145  & 4188        & 4141.5      & 4145.1   \\
		$\eta_{c2}$(2 $^1D_2$)& 4138&                 & 4158    & 4099  & 4150  & 4182        & 4157.6      & 4159.1    \\
		\noalign{\smallskip}\hline                                &
	\end{tabular}
\end{table}
\section{Decay properties}\label{sec3}
For quarkonium, most of the decay properties are dependent on the wave function. Hence to test the wavefunctions as obtained in the previous section, we calculate leptonic decay widths and radiative decay widths (M1 $\&$ E1) of some charmonium states.
\subsection{Leptonic decay widths}
The leptonic decay widths of the vector states are  calculated using the Van Royen-Weisskopf formula \cite{vanroyen,kwong}
\begin{equation*}
\Gamma_{ee} = \frac{4\alpha^2e_c^2}{M_{nS}}|R_{nS}(0)|^2\left( 1-\frac{16\alpha_s}{3\pi} \right)~,
\end{equation*}
where $M_{nS}$ is the mass for $nS$ state, $e_c$ is the charm quark charge in unit of electron charge, $\alpha$ is the fine
structure constant, $\alpha_s \approx \alpha_s(2m_c)$ is the strong coupling constant, $R_{ns}(0)$ is the radial $nS$ wave function at the origin. The terms in parenthesis are the QCD radiative correction factor. Obtained results are listed in Table \ref{lep}.
\begin{table}[h]
	\centering
	\caption{Average radii (in fm) and square of the radial wave function at the origin (in GeV$^3$)}
	\label{radii}       % Give a unique label
	%For LaTeX tables use
	\begin{tabular}{llllllll}
		\hline\noalign{\smallskip}
		State        & $\sqrt{<r^2>}$ & \cite{debastiani} & \cite{akbar1}  & \cite{qing} & $|R(0)|^2$ & \cite{debastiani} & \cite{akbar1}   \\
		\noalign{\smallskip}\hline\noalign{\smallskip}
		$\eta_c$(1S) & 0.380          & 0.375      & 0.3655 &      & 1.649      & 1.5405     & 1.2294  \\
		$\eta_c$(2S) & 0.863          & 0.839      & 0.8328 &      & 0.731      & 0.7541     & 0.8717  \\
		$\eta_c$(3S) & 1.250          & 1.210      & 1.2072 &      & 0.573      & 0.6088     & 0.683   \\
		$\eta_c$(4S) & 1.584          & 1.531      & 1.5306 &      & 0.502      & 0.5430     & 0.5994  \\
		$\eta_c$(5S) & 1.885          &            & 1.8225 &      & 0.461      &            & 0.5503  \\
		$J/\psi$     & 0.434          & 0.421      & 0.4143 & 0.41 & 0.976      & 1.1861     & 1.97675 \\
		$\psi$(2S)   & 0.897          & 0.867      & 0.8627 & 0.91 & 0.897      & 0.7092     & 0.7225  \\
		$\psi$(3S)   & 1.274          & 1.230      & 1.2287 & 1.38 & 1.274      & 0.5914     & 0.6006  \\
		$\psi$(4S)   & 1.603          & 1.547      & 1.5478 & 1.87 & 1.603      & 0.5340     & 0.5417  \\
		$\psi$(5S)   & 1.902          &            & 1.8370 & 2.39 & 1.902      &            & 0.50538 \\
		1 $^1P$      & 0.700          & 0.678      & 0.6738 &      & $\approx 0$        & 0          & $\approx 0$     \\
		1 $^3P$      & 0.712          & 0.689      & 0.7173 & 0.71 & $\approx 0$        & 0          & $\approx 0$     \\
		2 $^1P$      & 1.108          & 1.071      & 1.0697 &      & $\approx 0$        & 0          & $\approx 0$     \\
		2 $^3P$      & 1.120          & 1.082      & 1.1107 & 1.19 & $\approx 0$        & 0          & $\approx 0$     \\
		1 $^1D$      & 0.931          & 0.899      & 0.8984 &      & $\approx 0$        & 0          & $\approx 0$     \\
		1 $^3D$      & 0.932          & 0.901      & 0.9179 & 0.96 & $\approx 0$        & 0          & $\approx 0$     \\
		2 $^1D$      & 1.304          & 1.258      & 1.2595 &      & $\approx 0$        & 0          & $\approx 0$     \\
		2 $^3D$      & 1.305          & 1.261      & 1.1914 & 1.44 & $\approx 0$        & 0          & $\approx 0$    \\
		\noalign{\smallskip}\hline
	\end{tabular}
\end{table}
\subsection{M1 radiative transitions}
Magnetic dipole (M1) radiative transitions obey the selection rule $\Delta L = 0$ and $\Delta S = \pm 1$. The M1 widths are evaluated using the formula \cite{barnes}
\begin{equation*}
\Gamma_{M1}(n^{~2S+1}L_J \rightarrow n^{\prime ~2S^\prime+1}L'_{J'}) = \frac{4}{3} \frac{2J^\prime +1}{2L+1}\delta_{LL^\prime}\delta_{S,S^\prime +1} e_c^2 \frac{\alpha}{m_c^2} |<\psi_f|\psi_i>|^2 E_\gamma^3 \frac{E_f}{M_i} ~,
\end{equation*}
where $E_\gamma$ is the emitted photon energy, $<\psi_f|\psi_i>$ is the overlap integral involving initial and final radial wavefunctions, $E_f$ is the total energy of the final state and $M_i$ is the mass of the initial state. Calculated M1 widths are listed in Table \ref{m1}.
\begin{table}
	\centering
	\caption{Leptonic decay widths (in keV)}
	\label{lep}       % Give a unique label
	%For LaTeX tables use
	\begin{tabular}{lllllll}
		\hline\noalign{\smallskip}
		State      & Present & Exp \cite{pdg}  & \cite{akrai} & \cite{soni}  & \cite{akbar2}  & \cite{qing} \\
		\noalign{\smallskip}\hline\noalign{\smallskip}
		$J/\psi$   & 4.979   & 5.55 $\pm$ 0.14 $\pm$ 0.02 & 3.623 & 2.925 & 1.8532 & 6.60 \\
		$\psi$(2S) & 2.137   & 2.33 $\pm$ 0.04        & 1.085 & 1.533 & 0.5983 & 2.40 \\
		$\psi$(3S) & 1.460   & 0.86 $\pm$ 0.07        & 0.748 & 1.091 & 0.3812 & 1.42 \\
		$\psi$(4S) & 1.131   & 0.58 $\pm$ 0.07        & 0.599 & 0.856 & 0.2847 & 0.97 \\
		$\psi$(5S) & 0.930   &                            & 0.508 & 0.707 & 0.2286 & 0.70 \\
		\noalign{\smallskip}\hline
	\end{tabular}
\end{table}
\subsection{E1 radiative transitions}
Electric dipole (E1) radiative transitions obey the selection rule $\Delta L = \pm 1$ and $\Delta S = 0$. The E1 widths are evaluated using the formula \cite{barnes}
\begin{equation*}
\Gamma_{E1}(n^{~2S+1}L_J \rightarrow n^{\prime ~2S^\prime+1}L'_{J'}) = \frac{4}{3} C_{fi} \delta_{SS^\prime} e_c^2 \alpha |<\psi_f|r|\psi_i>|^2 E_\gamma^3 \frac{E_f}{M_i} ~,
\end{equation*}
where $<\psi_f|r|\psi_i>$ is the spatial matrix element involving the initial and final radial wavefunctions, and $C_{fi}$ is the angular matrix element given by
\begin{equation*}
C_{fi} = \text{max}(L,L')(2J'+1) \left\{ \begin{array}{c c c}
L'&J'&S\\
J&L&1\\
\end{array}\right\}^2
\end{equation*}
E1 widths obtained from the present analysis are listed in Table \ref{e1}.
\begin{table}
	\centering
	\caption{M1 radiative partial widths (in keV)}
	\label{m1}       % Give a unique label
	%For LaTeX tables use
	\begin{tabular}{lllllll}
		\hline\noalign{\smallskip}
		Transition  & Present & Exp \cite{pdg} & \cite{barnes} & \cite{soni}  & \cite{akrai} & \cite{deng} \\
		\noalign{\smallskip}\hline\noalign{\smallskip}
		$1~^3S_1 \rightarrow 1~^1S_0$ & 2.66    & 1.58 $\pm$ 0.37 & 2.9    & 2.722 & 1.647 & 2.39 \\
		$2~^3S_1 \rightarrow 2~^1S_0$ & 0.17    & 0.21 $\pm$ 0.15 & 0.21   & 1.172 & 0.135 & 0.19 \\
		$2~^3S_1 \rightarrow 1~^1S_0$ & 5.02    & 1.00 $\pm$ 0.15 & 4.6    & 7.506 & 69.57 & 7.80 \\	
		\noalign{\smallskip}\hline
	\end{tabular}
\end{table}
\begin{table}
	\centering
	\caption{E1 radiative partial widths (in keV)}
	\label{e1}       % Give a unique label
	%For LaTeX tables use
	\begin{tabular}{llllllll}
		\hline\noalign{\smallskip}
		Transition  & Present & Exp \cite{pdg}  & \cite{barnes} & \cite{soni} & \cite{akrai}  & \cite{deng} & \cite{qing} \\
		\noalign{\smallskip}\hline\noalign{\smallskip}
		$2~^3S_1 \rightarrow1~^3P_2$ & 29.78   & 27.99 $\pm$ 0.96   & 38     & 62.312  & 7.07   & 36   & 43   \\
		$2~^3S_1 \rightarrow1~^3P_1$ & 49.31   & 28.67 $\pm$ 1.05   & 54     & 43.292  & 10.39  & 45   & 62   \\
		$2~^3S_1 \rightarrow1~^3P_0$ & 49.38   & 28.78 $\pm$ 0.98   & 63     & 21.863  & 11.93  & 27   & 74   \\
		$1~^3P_2 \rightarrow1~^3S_1$ & 436.45  & 374.30 $\pm$ 19.73 & 424    & 157.225 & 233.85 & 327  & 473  \\
		$1~^3P_1 \rightarrow1~^3S_1$ & 319.08  & 288.12 $\pm$ 16.09 & 314    & 146.317 & 189.86 & 269  & 354  \\
		$1~^3P_0 \rightarrow1~^3S_1$ & 175.78  & 151.2 $\pm$ 9.99   & 152    & 112.030 & 118.29 & 141  & 167  \\
		$2~^3D_1 \rightarrow1~^3P_2$ & 6.07    & \textless{}17.4    & 4.9    & 5.722   & 6.45   & 5.4  & 5.8  \\
		$2~^3D_1 \rightarrow1~^3P_1$ & 159.05  & 67.73 $\pm$ 6.73   & 125    & 93.775  & 139.52 & 115  & 150  \\
		$2~^3D_1 \rightarrow1~^3P_0$ & 425.59  & 187.68 $\pm$ 17.72 & 403    & 161.504 & 343.87 & 243  & 486 \\	
		\noalign{\smallskip}\hline
	\end{tabular}
\end{table}
\section{Discussion and Summary}\label{sec4}
In the present work, we have numerically solved the Schr\"odinger equation for charmonium system using the DVR scheme of Colbert and Miller. The Hamiltonian matrix was constructed and diagonalised to obtain the masses and wavefunctions of charmonium states. In Table \ref{mass}, we compare the masses of radially and orbitally excited $c\bar{c}$ states with experiment \cite{pdg} and other theroretical predictions \cite{barnes,qing,akrai,soni,akbar1,debastiani}. Authors in refs. \cite{barnes,akrai,soni,akbar1,debastiani} have also used Cornell type potential to study the $c\bar{c}$ system, where as in ref. \cite{qing}, authors use a screened potential. From Table \ref{mass} we see that the predictions using the DVR method are in good agreement with experiment and other theoretical predictions. In Table \ref{radii} we have compared our predictions for the root mean square radii ($\sqrt{<r^2>}$) and the square of the radial wavefunction at the origin ($|R(0)|^2$) with other theoretical predictions \cite{qing,akbar1,debastiani}. The values of radial wavefunctions at the origin are important inputs for calculating quarkonium production cross-sections \cite{origin} and various decay amplitudes. We present our results for leptonic decays in Table \ref{lep} in comparison with experiment and other models. Our predictions for lower states are in good agreement with the experimental results. For higher excited states, our predictions are higher than the experimental results. We present results of E1 and M1 radiative transitions in Tables \ref{m1} \& \ref{e1} respectively. Radiative transitions in quarkonia are important because they are one of the few mechanisms that produce transitions among $q\bar{q}$ states with different quantum numbers. This decay mechanism also help to produce excited P-wave states and F-wave states which are otherwise difficult to achieve \cite{barnes}. M1 decays in particular allows to access spin-singlet states. From Tables \ref{m1} \& \ref{e1}, we see that there is a wide range of prections for the radiative decay widths even though all these models \cite{barnes,soni,akrai,deng} employs a Cornell type potential. This may be due to the difference in wavefunctions of charmonium states used in these models. Our predictions for radiative decays are in accordance with experiment and other theoretical predictions. Inclusion of higher multipole contributions, coupled channel effects, relativistic corrections, etc. would give a better fit to the experimental results.
\\
In summary, in this article we have successfully employed the DVR  method to investigate the spectra and decays of charmonium. The obtained results of present study are in good agreement with experimental data and with other theoretical models.

\bibliographystyle{phaip}
\bibliography{manuscript_bha.bib}
\end{document}